# Mechanical performance of hybrid polymer-lipid vesicles with leaflet asymmetry engineered using microfluidics

Yuting Huang[1], Arash Manafirad[2], Simon Matoori[1,3], Laura R. Arriaga[4], Sijie Sun[1], Anqi Chen[1], Anthony D. Dinsmore[2], David J. Mooney[1,5], and David A. Weitz[1,5*]

1. John A. Paulson School of Engineering and Applied Sciences Harvard University, Cambridge, MA, 02138, USA
2. Department of Physics, University of Massachusetts Amherst, Amherst, MA, 01003, USA
3. Faculté de Pharmacie, Université de Montréal, Montreal, Quebec H3T 1J4, Canada
4. Department of Theoretical Condensed Matter Physics, Condensed Matter Physics Center and Instituto Nicolás Cabrera, Universidad Autónoma de Madrid, 28049, Madrid, Spain
5. Wyss Institute for Biologically Inspired Engineering, Harvard University, Boston, MA, 02115, USA.

Corresponding author email*: weitz@seas.harvard.edu

**Abstract:**

Lipid vesicles consist of aqueous cores surrounded by a bilayer of phospholipids. Hybrid polymer-lipid vesicles incorporate both polymers and lipids, offering promising properties for developing pharmaceuticals, biosensors, and artificial cells. The hybrid vesicles can be symmetric, in which their two leaflets contain identical compositions, or asymmetric, in which the leaflets possess dissimilar compositions and can lead to dramatically modified properties. However, methods to produce both symmetric and asymmetric hybrid vesicles result in heterogenous compositions and sizes, making it challenging to quantify the effect of asymmetry and limiting applications. Here, we use a microfluidic approach to produce hybrid vesicles containing symmetric or asymmetric leaflets with precisely engineered compositions. We find the vesicles with asymmetric leaflets are significantly stiffer and tougher than those with symmetric leaflets; moreover, the lateral diffusivity of lipids is greatly decreased. The structure for improved toughness consists of an inner leaflet that is a stretchable lipid leaflet and an outer leaflet that is a fully continuous polymer leaflet. This technique of precisely engineering asymmetric structures may be applied to hybrid vesicles composed of block copolymers and phospholipids dissolvable in chloroform and hexane, further expanding their applications.

**Significance Statement**

Vesicles, which consist of aqueous cores surrounded by lipid bilayers, are widely explored for drug encapsulation and delivery due to their resemblance to cell membranes. Polymersomes, formed from bilayers of block copolymers, offer enhanced toughness through their synthetic and expanded material properties but lack the biocompatibility of lipid vesicles. Hybrid polymer-lipid vesicles combine the biocompatibility of lipids with the chemical versatility of polymers. Here, we report a novel method for producing asymmetric vesicles with distinct lipid and polymer leaflets that offer even greater advantages. These vesicles are nearly as stiff yet much tougher compared to polymersomes, while maintaining biocompatibility thanks to their lipid based inner leaflet.

**Introduction**

Lipid vesicles consist of aqueous cores surrounded by a bilayer of phospholipids; they are intrinsically biocompatible and widely used as delivery vehicles for pharmaceuticals, food, and cosmetics (1-8). However, lipid vesicles suffer from poor mechanical stability and limited chemical functionality (9-12). By comparison, polymersomes are synthetic analogs of lipid vesicles made from much larger amphiphilic polymers(13, 14). Polymersomes can exhibit properties differing by one or multiple orders of magnitude, greatly extending the application of lipid vesicles (15-17). Despite the improved mechanical robustness, polymersomes often lack sufficient biocompatibility, limiting their usage in biotechnology (9-12, 18). Hybrid polymer-lipid vesicles consist of a bilayer of polymer and lipid mixtures; they have the potential advantage of combining the biocompatibility of lipid vesicles with the chemical versatility of polymersomes (9-11, 19-22). For example, when a polymer such as PEG is presented on the outer surface of the vesicles, the vesicles can evade the immune systems in a 'stealthy' mode, reducing immune clearance, while the lipid component facilitates fusion with target cells. These hybrid vesicles offer unique advantages in drug delivery. Their properties, however, are not understood due to the difficulty in assembling and assessing them. With drastically different chemistry and size, the polymers and lipids can arrange into different microstructures, which lead to an array of properties unattainable in lipid vesicles and pure polymersomes(10, 11, 23). Depending on the type of polymer and the volume ratio between polymers and lipids, the hybrid vesicles can possess properties either in between or exceeding the range exhibited by lipid vesicles and polymersomes(19, 24-26). The vesicles can possess symmetric membranes, where the polymers and lipids are identical in each monolayer. The polymers and lipids can homogeneously mix when the entropy dominates over the chemical potential, or phase separate when the potential energy gain of assembly outweighs the entropy gain(22, 23, 27-30). The vesicles can possess asymmetric membranes, where the two monolayers contains dissimilar compositions(6, 31-36). The individual structure of each monolayer, together with their coupling, determines the property of the vesicle. The asymmetric structure introduces a new degree of complexity to vesicles (37-50). Conventional methods, such as rehydration and electroformation, struggle to control the size, composition, and structure of vesicles(10, 28, 35). How these intricate microstructures dictate the macroscopic properties remains elusive. Thus, the absence of both a controlled assembly method and a thorough study of vesicle properties hinders the potential for many applications.

In this work, we produce the hybrid vesicles with controlled membranes, both symmetric and asymmetric, and analyze their mechanical properties. We generate the vesicles using a novel multiple emulsions approach, made with microfluidics, which enables engineering each leaflet to have precise compositions. Symmetric vesicles are made from water-in-oil-in-water double emulsions, while asymmetric vesicles are made from triple emulsions, with a water core surrounded by two oil shells. Using micropipette aspiration, we find that asymmetric hybrid vesicles possess an enhanced stretching modulus and toughness as compared to symmetric hybrid vesicles. Using Fluorescence Recovery After Photobleaching (FRAP), we determine that the lipid diffusion coefficient is lower in asymmetric hybrid vesicles as compared to that in symmetric hybrid vesicles, even when the amounts of polymer are similar in the vesicles. Membrane permeability test further shows reduced permeability in asymmetric vesicles as compared to symmetric vesicles.

## Results and Discussion
### Microfluidic Production of W/O/W Double Emulsion Templated Symmetric Hybrid Vesicles and Their Fluorescence Characterization

To fabricate symmetric polymer-lipid vesicles, we use a glass capillary microfluidics device to generate water-in-oil-in-water (W/O/W) double emulsions as templates for vesicles(51, 52). During operation, we introduce an aqueous phase containing 10 wt% poly(ethylene glycol) (PEG) through the innermost capillary from the left-hand side at a flow rate of 300 µl/hr. This PEG solution becomes the interior of the vesicles. Also, this PEG solution significantly enhances the optical contrast between the double emulsion cores and the outer aqueous media. Simultaneously, we inject an oil phase containing 80 wt% 1,2-dioleyl-sn-glycero-3-phosphocholine (DOPC) and 20 wt% 5 kDa poly(ethylene glycol)-block-10 kDa poly(D,L-lactic acid) (PEG-b-PLA) , dissolved at 5 mg/ml in a mixture of 36 vol% chloroform and 64 vol% hexane, through the left tapered capillary at the same flow rate. The hydrophobic coating on the left capillary allows large water-in-oil emulsions to form inside the capillary, as shown in Figure 1A. Next, we inject 10 wt% poly(vinyl alcohol) (PVA) through the right tapered capillary at a flow rate of 3000 µl/hr. The hydrophilic coating on the right capillary prevents the middle oil phase from wetting its surface.

Under these conditions, W/O/W double emulsions with an outer diameter of 80 µm form at the junction between the right and left capillaries, as depicted in the bright-field image in Figure 1A (bottom). We collect these emulsions in a solution composed of 50 vol% PBS buffer and 50 vol% water, ensuring the same osmolarity as the inner phase. To assess the uniformity of the vesicles, we capture optical images and analyze their diameters using a contour detection algorithm developed in MATLAB. Our analysis reveals that the size of the collected double emulsions exhibit a coefficient of variation (CV) of approximately 5.2% (Figure 1B). This level of uniformity is consistent with the range reported for vesicles produced by other microfluidic techniques (53, 54). The PEG-b-PLA and DOPC, immersed in the middle oil shell, adsorb to the interfaces between the aqueous phase and the oil phase; as the oil leaves the membrane, the two interfaces come into contact to form polymer-lipid vesicles with symmetric leaflets, as illustrated by Figure 1C. This oil detachment is called the dewetting phenomenon. During dewetting, two monolayers come into close contact and interact via hydrophobic interactions to form a bilayer, which results in the expulsion of excess oil as small budded droplets from the membrane (55-62). When viewed under the microscope, the oil shells in double emulsions appear to become thinner and exit the membrane in the form of oil bubbles (66, 67), allowing the double emulsions to form vesicles, as shown by the bright field images in Figure 1D. Importantly, DOPC is a fluid-phase lipid at room temperature, whereas PEG-b-PLA is a solid-phase polymer(63, 64). This combination of fluid-like and solid-like components introduces a new level of complexity in the membranes, akin to the complexity found in cell membranes(65, 66). Furthermore, PEG-b-PLA is more than ten times larger than DOPC, resembling the way small molecules are embedded with large proteins in cell membranes. This size disparity allows us to explore how significant differences in molecular sizes influence the overall properties of the membrane. To characterize the distribution of polymers and lipids within the membrane, we incorporate 3 wt% of the lipophilic dye 1,2-dioleoyl-sn-glycero-3-phosphoethanolamine-N-(lissamine rhodamine B sulfonyl) (18:1 Liss Rhod PE) into the lipid leaflet, and 3 wt% of the block copolymer dye fluorescein isothiocyanate-PEG-b-PLA (FITC-PEG 5000 Da-b-PLA 10000 Da) into the polymer leaflet. We then examine the vesicles using confocal microscopy with a 10X objective. This method reveals a uniform distribution of both dyes across the vesicle surface, as shown in Figure 1E. The lipids are thus embedded in between polymers throughout the membrane. The fluid phase lipid DOPC likely facilitates the distribution of the solid phase PEG-b-PLA across the vesicle surface. Using confocal microscope with a10X objective, we cannot detect any phase separation of polymers and lipids; however, separate polymer and lipid domains exist below the resolution of the objective. To investigate how the solid phase of lipids affects the overall membrane, we incorporate 40 wt% DPPC, a solid-phase lipid at room temperature, into the membrane fabrication. The resulting membrane is comprised of 40 wt% DPPC, 40 wt% DOPC, and 20 wt% PEG-b-PLA, as shown in Figure 1F. Additionally, we include 3 wt% naphthopyrene, which associates with DPPC-rich domains, and 3 wt% 18:1 Liss Rhod PE, which associates with DOPC-rich domains. Using confocal microscopy, we observe a phase-separated membrane, where red fluorescence indicates DOPC-rich domains, blue fluorescence highlights DPPC-rich domains, and non-fluorescent regions represent polymer-rich domains, as depicted in Figure 1F. These macroscopic phase-separated regions form after 24 hours of collection, during which smaller scattered phase-separated regions coalesce into one polymer rich region and one lipid rich region, as shown in the supplement Figure S1. These observations lead us to conclude that solid lipids and solid polymers phase-separate more readily than liquid lipids and solid polymers. Liquid lipids likely promote uniform distribution of molecules through faster diffusion. This study demonstrates a potential microfluidic approach for creating symmetric polymer-lipid vesicles. Similar microfluidic approach has been used to create various types of pure polymersomes and pure lipid vesicles, suggesting the microfluidics platform can be applicable to various formulations.(55-62, 67-70)

**Microfluidic Production of W/O1/O2/W Triple Emulsion Templated Asymmetric Hybrid Vesicles and Their Fluorescence Characterization**

To make asymmetric hybrid vesicles, we employ a novel approach using the same device that produces symmetric hybrid vesicles but modify the procedure to produce water-in-oil-in-oil-in-water (W/O/O/W) triple emulsions. In addition to one oil phase, a second oil phase containing PEG-b-PLA at the same concentration is injected through the gap between the left cylindrical capillary and the outermost square capillary, also at a flow rate of 300 μl/hr. Under these conditions, W/O/O/W triple emulsions form between the tapered capillaries, as shown by the bright field image in Figure 2A. Our analysis reveals that the collected triple emulsions exhibit a coefficient of variation (CV) of approximately 5.5%, as shown in Figure 2B. This level of uniformity is consistent with the range reported for vesicles produced by other microfluidic techniques. The PEG-b-PLA in the outer oil shell predominantly diffuses to the outer water-oil interface, while DOPC in the inner shell migrates mainly to the inner water-oil interface. As the oil exits the membrane, the two interfaces come into contact, resulting in the formation of a polymer-lipid vesicle with asymmetric leaflets. This process forces the oil to dewet from the membrane, as illustrated in Figure 2C. Under the 10x objective, we observe the oil exiting the triple emulsions by forming oil caps at a contact angle to the membrane, as depicted by the bright-field images in Figure 2D. Moreover, to investigate the molecular distribution within the membrane, we incorporate 3 wt% of a red fluorescent lipid, 18:1 Liss Rhod PE, and 3 wt% of a green fluorescent polymer, FITC-PEG-b-PLA, into the membrane. Using confocal microscopy, we observe a uniform distribution of both the lipid and polymer across the membrane, as indicated by the consistent red and green fluorescence shown in Figure 2E. This result suggests that the lipids and polymers are distributed throughout the membrane, likely in separate leaflets.

To characterize bilayer asymmetry, we measure the degree of asymmetry, defined as the percentage of molecules that remain asymmetrically distributed in the bilayer (6, 71-73). For vesicles with polymer inner and lipid outer leaflets, labeled as formulation F4, we incorporate 3 wt% of fluorescent lipids, 1,2-dioleoyl-sn-glycero-3-phosphoethanolamine-N-(7-nitro-2-1,3-benzoxadiazol-4-yl) (ammonium salt) (DOPE-NBD), into the lipid outer leaflet during fabrication. We then add a 1M dithionite solution, a quencher for DOPE-NBD, and observe the fluorescence under confocal microscopy five minutes post-addition. The fluorescence signal of the vesicles drops to 34%, as shown by the right images in Figure 2F. Further addition of a salt solution causes the vesicles to break, completely extinguishing the fluorescence. These results suggest that 34% of DOPE-NBD resides in the inner leaflet, while 66% remains in the outer leaflet, resulting in a degree of asymmetry of 66% for these vesicles. To investigate the degree of asymmetry in vesicles with lipid inner and polymer outer leaflets, labeled as formulation F5, we switch the two oils during production and incorporate 3 wt% of DOPE-NBD into the inner DOPC leaflet. After adding a 1M dithionite solution, we observe that the fluorescence signal of these vesicles drops to 78%, as shown by the left images in Figure 2G. Subsequent addition of a salt solution with higher osmotic pressure than the vesicle cores cause the vesicles to break, fully extinguishing the fluorescence. These findings indicate that 78% of DOPE-

NBD is in the inner leaflet, while 22% is in the outer leaflet, resulting in a degree of asymmetry of 78% for these vesicles. The higher degree of asymmetry in formulation F5 compared to F4 could be attributed to the increased polymer introduction from the thick outer oil shell, leading to fewer lipids being incorporated into the membrane and thus higher asymmetry.  Strategies to improve membrane asymmetry, such as increasing the polymer phase thickness or concentration in the inner phase and reducing the lipid phase thickness or concentration in the outer phase, have been suggested in prior simulation studies(36). As noted in that work, we can increase the asymmetry of the vesicle membrane by controlling the relative thicknesses of each layer and by tuning the concentration imbalance across the leaflets. While we did not explore these strategies in the current study due to scope limitations, we have added discussions for improving the degree of asymmetry.  Moreover, the degree of asymmetry is currently defined based on the asymmetric lipid distribution without determining the polymer distribution. To more accurately determine the composition in each leaflet as reported in literature(74), other non-fluorescence method would be needed for future work. We suggest a future direction of this work as to determine the ratio of total amounts of lipids and polymers incorporated or the exact composition in each leaflet using both fluorescent and nonfluorescent methods in comparison.

**Mechanical Properties of the Vesicles Assessed by Micropipette Aspiration Measurements**

To study the mechanical properties of the vesicles, we conduct micropipette aspiration measurements (30, 75-85). To obtain the stretching modulus, our vesicles are slightly inflated, with an inner osmolarity that is 20 mOsmo higher than that of the outer aqueous media. The procedure is described in detail in the supplementary document. We plot the tension versus strain, where the slope represents the stretching modulus and the area under the curve represents the toughness.

To investigate the effect of composition on the mechanical properties of the vesicles, we examine their tension versus strain curves. As a reference, we consider the DOPC vesicles, labeled F1. We find that the tension versus strain relationship is linear, as evidenced by the light blue colored curve in Figure 3C. This behavior is typical for fluid vesicles in the areal expansion regime. However, the stretching modulus, derived from the slope of the tension as a function of strain, is lower than that of DOPC vesicles formed by electroformation. We hypothesize that the process of vesicle formation from emulsion templates may result in different packing of the monolayer as compared to processes such as electroformation and rehydration. Processes such as electroformation and rehydration are known to influence the structure of the vesicles differently. It is common for any process to impact the resultant vesicles. Like the other processes, the samples prepared from one process can be internally compared for their compositions.  As we increase the polymer fraction in vesicles to 20 wt%, the data exhibit a sharper initial linear rise. When we increase the polymer fraction to 80 wt%, there is a threshold tension at which strain is observed. This behavior suggests that the vesicle with 80 wt% polymer, after the initial yielding, remains a stiff solid consisting of large polymer networks. We also measure vesicles with polymer on the inside and lipid on the outside, labelled as F4, which possess 66% asymmetry of the lipids. The slope of the initial rise is greater than that of the DOPC vesicles. Finally, we measure the asymmetric vesicles with polymer outside and lipid inside. The data exhibit a sharp rise at zero strain and a significantly steeper slope, as shown by the red curve in Figure 3C. Notably, we find the the stretching modulus to be approximately two to three times as large as the vesicle with a higher degree of asymmetry, as shown by the dark blue curves in Figure 3C.

To confirm the fluid or solid state of the vesicles, we examine optical images of vesicles at the moment of rupture, as shown in Figure 3D. Vesicles composed of DOPC lipids, symmetric hybrid vesicles composed of 20 wt% polymers, and asymmetric vesicles with 66% asymmetry all display smooth membranes upon breaking, consistent with both leaflets being fluid-like, as depicted by F1, F2, and F4 in Figure 3D. In contrast, symmetric hybrid vesicles with 80 wt% polymer, asymmetric hybrid vesicles with 78% asymmetry, and polymersomes all exhibit wrinkles upon breaking. Interestingly, we also find that the highly asymmetric vesicles, labeled as F6, exhibit an order of magnitude higher stretching modulus than all symmetric hybrid vesicles, labelled as F1 to F3, while possessing several times higher of lysis strain and toughness than polymersomes, labelled as F5, as shown by Figure 3E. Notably, the symmetric vesicle shows higher polymer fluorescence and lower lipid fluorescence, whereas in asymmetric vesicles, the vesicles exhibit a more balanced polymer and lipid fluorescence.  This result suggests that the asymmetric vesicles are tougher than the symmetric vesicles, likely due to the asymmetric structure and not just the polymer content.

## Lipid Diffusion in Vesicle Membrane Assessed by Fluorescence Recovery after Photobleaching (FRAP) Measurements

To investigate the presence of domains exist in symmetric versus asymmetric vesicles, we employ Fluorescence Recovery After Photobleaching (FRAP) to determine lipid mobilities (86, 87). We label the membranes with 2 mol% of the lipophilic fluorescent dye 1,2-dioleoyl-sn-glycero-3-phosphoethanolamine-N-(carboxyfluorescein) (ammonium salt) (DOPE-carboxy fluorescein). To investigate how composition and structure impact the fluidity of the overall membrane, we apply FRAP to a range of formulations, labeled F1 to F5, as listed in the table in Figure 4B. The normalized fluorescence intensity in a DOPC vesicle recovers to a plateau intensity, $I_p$, which is 100% of the unbleached control fluorescence, as shown by the blue curve in Figure 4C.

To determine whether the lipids are diffusing normally, we employ a semi-log plot to obtain $\log\left(I_p - I(t)\right)$ versus $t$ for vesicles of different compositions, as shown in Figure 4D. We observe linear behaviors for all vesicles, indicating an exponential recovery consistent with normal diffusive behaviors for the lipids. Thus, we fit the normalized recovery fluorescence to $I(t) = I_o(1 - \exp\left(-\frac{t}{\tau_D}\right))$ (88). Using the intensity recovery plots in Figure 4C, we determine the half time, $\tau_{1/2} = -(\ln 0.5)\tau_D$, defined as the time it takes to recover to half of the plateau intensity. We repeat the FRAP experiment with varying bleached areas, with bleached radius $r$, as illustrated by the schematic for a DOPC vesicle in Figure 4E, and obtain their half time of recovery dependent on bleached size. We plot the bleached area size versus recovery half time for each vesicle type. We observe linear relationships between $r^2$ and $\tau_{1/2}$ for all vesicle types, as shown by Figure 4F. As the polymer content increases, we observe a similar prolonged recovery time, as shown by Figure 4C. As we increase the degree of asymmetry from F4 to F5, we find that the slope decreases significantly. This suggests that while increased lipid content in one leaflet may enhance local diffusion, the presence of a dense, continuous polymer leaflet can dominate the overall diffusional behavior. The net effect is a reduction in lipid diffusion in asymmetric vesicles compared to symmetric lipid vesicles.

To quantify the diffusion coefficients, $D$, we use the equation $D = \frac{r^2}{4\tau_{1/2}}$ (88), derived from the slope of the $r^2$ versus $4\tau_{1/2}$ line and summarize the statistics in Figure 4G (89). We plot the mean and standard deviation of the diffusion coefficients for all types of vesicles, as shown in Figure 4G. For DOPC vesicles, we determine the lipid diffusion coefficient to be approximately 6.1 $\mu m^2/s$, as shown by the light blue bar in Figure 4G. This diffusion coefficient of DOPC vesicles is within the range reported in literature (67-70). As polymer content increase to 20 wt% and 80 wt%, the diffusion coefficient drops approximately by 33% to 4.0 $\mu m^2/s$ and by 78% to 1.3 $\mu m^2/s$, as shown by the purple and green bars, respectively, in Figure 4G. For asymmetric vesicles with 66% asymmetry, the diffusion coefficient is 0.9 $\mu m^2/s$, which is slightly lower than that of the vesicles with 80 wt% polymers, as shown by the yellow bar in Figure 4G. By contrast, for vesicles with 78% asymmetry, the diffusion coefficient further decreases by another factor of 3 to approximately 0.2 $\mu m^2/s$, as indicated by the red bar in Figure 4G. This decrease in diffusion coefficient suggests that the polymers form a fully continuous leaflet in these highly asymmetric vesicles, drastically slowing the lipid diffusion.

## Conclusion

In this work, we present a novel method to fabricate hybrid vesicles with engineered leaflet compositions and asymmetry. Our results show that a fully continuous polymer leaflet makes asymmetric hybrid vesicles significantly stiffer and tougher, with reduced fluidity and membrane pore size, as compared with symmetric hybrid vesicles. Leaflet asymmetry introduces a new degree of freedom to manipulate hybrid vesicles properties, expanding their potential applications in pharmaceuticals, biosensors, and artificial cells(12, 90-95), where precise control of vesicle properties is crucial.

## Materials and Methods

**Chemicals.** All lipids are purchased from Avanti Polar Lipids Inc. The polymer, 5kDa Poly(ethylene glycol)-block-10kDa poly(D,L-lactic acid) (PEG-b-PLA) and fluorescein isothiocyanate-5kDa PEG-b-10kDa PLA (FITC-5kDa PEG-b-10kDa PLA) are purchased from Polysciences Inc and Nanosoftpolymers Inc respectively.

**Micropipette Aspiration.** The pipette has a radius of $R_p$ and a suction pressure $\Delta P$, as shown by the schematic and images in the first row in Figure 3A (96). A tensile stress is provided on the vesicle membrane. Then, the vesicle has a projected length of $L$ inside the micropipette and a radius of $R_s$ outside of the micropipette, as shown by Figure 3A. We determine the areal strain, $\alpha$, by calculating surface area of the vesicle from its contour, as detailed in the supplement. Since the pressure is uniform in the interior of the vesicle, the membrane curvatures enable us to determine the membrane tension $\tau$ using the Laplace equation, $\tau = \frac{\Delta P}{2\left(\frac{1}{R_p} - \frac{1}{R_s}\right)}$ (85).

**FRAP.** We photo-bleach the fluorescent lipids within a circular region. We observe the gradual recovery of fluorescence in the bleached area. We measure the recovered fluorescence intensity in the bleached area, $I_b(t)$, over time *t* by averaging the pixel intensities. We also track the fluorescence intensity in a control area in the unbleached region, $I_c(t)$, with the same radius as the bleached circular disk. We determine the normalized fluorescence intensity $I(t) = \frac{I_b(t)}{I_c(t)}$. We use the FRAP module using a 5X or 10X objective. The bleached radius is determined by drawing a line across the bleached region and identifying the pixel locations where the intensity gradient reaches local maxima. These points correspond to the edges of the bleached circle. The measured intensities are normalized to the prebleach intensities of the region of interest (ROI). We observe the fluorescence intensity virtually recovers to the prebleached intensity. Thus, we simplify the normalized intensity recovery fitting to $I(t) = A \times (1 - \exp(-\frac{t}{\tau_D}))$. Supplementary video V9 of an asymmetric vesicle during FRAP experiment is available.

**Microfluidic Production.** A range of chemical compositions can be used to generate symmetric and asymmetric hybrid vesicles with a microfluidic device. Symmetric hybrid vesicles can be prepared using total concentrations ranging from 5–10 mg/ml, with a fixed lipid-to-polymer ratio. The chloroform-to-hexane volume ratio can vary between 1:1.3 and 1:1.8 for generating symmetric hybrid vesicles. Asymmetric hybrid vesicles are created by using one oil phase containing lipids and another oil phase containing polymers at the same concentration of lipids. For asymmetric vesicles with polymer in the inner leaflet and lipids in the outer leaflet, the inner oil phase contains 30 mg/ml of polymer, while the outer oil phase contains 30 mg/ml of lipids. Importantly, we used the same fabrication conditions and post-processing protocols for oil removal across all vesicle types. This consistency minimizes batch-to-batch variability and ensures that the effect of residual oil on the vesicle properties, if any, would be comparable across samples.

**Storage of Vesicles and Management of Solvent Evaporation.** We collect approximately 100-200 $ul$ of vesicles in 30-40ml of diluted PBS buffer in each vial. All vesicles are stored in open glass vials containing PBS buffer diluted with water to match the osmolarity of the vesicle inner cores. The vesicles are stored at 4°C, and the buffer is replaced with fresh buffer 2 to 3 times within the first 48 hours to facilitate the evaporation of the organic solvents.

**Supplement**

Supplementary files and videos are available.

**Acknowledgement**

This work is based on support by the Harvard MRSEC funding DMR-2011754 and the Health@InnoHK program of the Innovation and Technology Commission of the Hong Kong SAR Government. We thank our funding sources for enabling us to do these experiments.

1. Chacko IA, Ghate VM, Dsouza L, Lewis SA. Lipid vesicles: A versatile drug delivery platform for dermal and transdermal applications. Colloids and Surfaces B: Biointerfaces. 2020;195:111262.
2. Jain S, Jain V, Mahajan S. Lipid based vesicular drug delivery systems. Advances in Pharmaceutics. 2014;2014(1):574673.
3. van der Meel R, Fens MH, Vader P, Van Solinge WW, Eniola-Adefeso O, Schiffelers RM. Extracellular vesicles as drug delivery systems: lessons from the liposome field. Journal of controlled release. 2014;195:72-85.
4. Zhang L, Chan JM, Gu FX, Rhee J-W, Wang AZ, Radovic-Moreno AF, et al. Self-assembled lipid− polymer hybrid nanoparticles: a robust drug delivery platform. ACS nano. 2008;2(8):1696-702.
5. Kauscher U, Holme MN, Björnmalm M, Stevens MM. Physical stimuli-responsive vesicles in drug delivery: Beyond liposomes and polymersomes. Advanced drug delivery reviews. 2019;138:259-75.
6. Pautot S, Frisken BJ, Weitz D. Engineering asymmetric vesicles. Proceedings of the National Academy of Sciences. 2003;100(19):10718-21.
7. Miatmoko A, Ayunin Q, Soeratri W. Ultradeformable vesicles: concepts and applications relating to the delivery of skin cosmetics. Therapeutic Delivery. 2021;12(10):739-56.
8. Reiner AT, Somoza V. Extracellular vesicles as vehicles for the delivery of food bioactives. Journal of agricultural and food chemistry. 2019;67(8):2113-9.
9. Kang JY, Choi I, Seo M, Lee JY, Hong S, Gong G, et al. Enhancing membrane modulus of giant unilamellar lipid vesicles by lateral co-assembly of amphiphilic triblock copolymers. Journal of colloid and interface science. 2020;561:318-26.
10. Chemin M, Brun P-M, Lecommandoux S, Sandre O, Le Meins J-F. Hybrid polymer/lipid vesicles: fine control of the lipid and polymer distribution in the binary membrane. Soft Matter. 2012;8(10):2867-74.
11. Le Meins J-F, Schatz C, Lecommandoux S, Sandre O. Hybrid polymer/lipid vesicles: state of the art and future perspectives. Materials today. 2013;16(10):397-402.
12. Go YK, Leal C. Polymer–lipid hybrid materials. Chemical reviews. 2021;121(22):13996-4030.
13. Discher DE, Ahmed F. Polymersomes. Annu Rev Biomed Eng. 2006;8(1):323-41.
14. Rideau E, Dimova R, Schwille P, Wurm FR, Landfester K. Liposomes and polymersomes: a comparative review towards cell mimicking. Chemical society reviews. 2018;47(23):8572-610.
15. Discher BM, Won Y-Y, Ege DS, Lee JC, Bates FS, Discher DE, et al. Polymersomes: tough vesicles made from diblock copolymers. Science. 1999;284(5417):1143-6.
16. Chang H-Y, Sheng Y-J, Tsao H-K. Structural and mechanical characteristics of polymersomes. Soft Matter. 2014;10(34):6373-81.
17. Discher BM, Bermudez H, Hammer DA, Discher DE, Won Y-Y, Bates FS. Cross-linked polymersome membranes: vesicles with broadly adjustable properties. The Journal of Physical Chemistry B. 2002;106(11):2848-54.
18. Matoori S, Leroux J-C. Twenty-five years of polymersomes: lost in translation? Materials Horizons. 2020;7(5):1297-309.
19. Schulz M, Binder WH. Mixed hybrid lipid/polymer vesicles as a novel membrane platform. Macromolecular rapid communications. 2015;36(23):2031-41.
20. Schulz M, Glatte D, Meister A, Scholtysek P, Kerth A, Blume A, et al. Hybrid lipid/polymer giant unilamellar vesicles: effects of incorporated biocompatible PIB–PEO block copolymers on vesicle properties. Soft Matter. 2011;7(18):8100-10.
21. Meyer CE, Abram S-L, Craciun I, Palivan CG. Biomolecule–polymer hybrid compartments: combining the best of both worlds. Physical Chemistry Chemical Physics. 2020;22(20):11197-218.
22. Hu S-W, Huang C-Y, Tsao H-K, Sheng Y-J. Hybrid membranes of lipids and diblock copolymers: From homogeneity to rafts to phase separation. Physical Review E. 2019;99(1):012403.
23. Magnani C, Montis C, Mangiapia G, Mingotaud A-F, Mingotaud C, Roux C, et al. Hybrid vesicles from lipids and block copolymers: Phase behavior from the micro-to the nano-scale. Colloids and Surfaces B: Biointerfaces. 2018;168:18-28.
24. Müller WA, Beales PA, Muniz AR, Jeuken LJ. Unraveling the Phase Behavior, Mechanical Stability, and Protein Reconstitution Properties of Polymer–Lipid Hybrid Vesicles. Biomacromolecules. 2023;24(9):4156-69.
25. Fauquignon M, Ibarboure E, Le Meins J-F. Membrane reinforcement in giant hybrid polymer lipid vesicles achieved by controlling the polymer architecture. Soft Matter. 2021;17(1):83-9.
26. Winzen S, Bernhardt M, Schaeffel D, Koch A, Kappl M, Koynov K, et al. Submicron hybrid vesicles consisting of polymer–lipid and polymer–cholesterol blends. Soft Matter. 2013;9(25):5883-90.

27. Nam J, Beales PA, Vanderlick TK. Giant phospholipid/block copolymer hybrid vesicles: Mixing behavior and domain formation. Langmuir. 2011;27(1):1-6.
28. Chen D, Santore MM. Hybrid copolymer–phospholipid vesicles: phase separation resembling mixed phospholipid lamellae, but with mechanical stability and control. Soft Matter. 2015;11(13):2617-26.
29. Huang C, Quinn D, Sadovsky Y, Suresh S, Hsia KJ. Formation and size distribution of self-assembled vesicles. Proceedings of the National Academy of Sciences. 2017;114(11):2910-5.
30. Fauquignon M, Ibarboure E, Le Meins J-F. Hybrid polymer/lipid vesicles: Influence of polymer architecture and molar mass on line tension. Biophysical Journal. 2022;121(1):61-7.
31. Richmond DL, Schmid EM, Martens S, Stachowiak JC, Liska N, Fletcher DA. Forming giant vesicles with controlled membrane composition, asymmetry, and contents. Proceedings of the National Academy of Sciences. 2011;108(23):9431-6.
32. Krompers M, Heerklotz H. A guide to your desired lipid-asymmetric vesicles. Membranes. 2023;13(3):267.
33. Peyret A, Zhao H, Lecommandoux S. Preparation and properties of asymmetric synthetic membranes based on lipid and polymer self-assembly. Langmuir. 2018;34(11):3376-85.
34. Tsai H-C, Yang Y-L, Sheng Y-J, Tsao H-K. Formation of asymmetric and symmetric hybrid membranes of lipids and triblock copolymers. Polymers. 2020;12(3):639.
35. Peyret A, Ibarboure E, Le Meins JF, Lecommandoux S. Asymmetric hybrid polymer–lipid giant vesicles as cell membrane mimics. Advanced Science. 2018;5(1):1700453.
36. Arriaga LR, Huang Y, Kim S-H, Aragones JL, Ziblat R, Koehler SA, et al. Single-step assembly of asymmetric vesicles. Lab on a Chip. 2019;19(5):749-56.
37. Wen L, Xiao K, Sainath AVS, Komura M, Kong XY, Xie G, et al. Engineered asymmetric composite membranes with rectifying properties. Advanced Materials. 2016;28(4):757-63.
38. Yu H, Qiu X, Moreno N, Ma Z, Calo VM, Nunes SP, et al. Self-assembled asymmetric block copolymer membranes: bridging the gap from ultra-to nanofiltration. Angewandte Chemie. 2015;127(47):14143-7.
39. Pabst G, Keller S. Exploring membrane asymmetry and its effects on membrane proteins. Trends in Biochemical Sciences. 2024.
40. Zhang Z, Kong X-Y, Xiao K, Liu Q, Xie G, Li P, et al. Engineered asymmetric heterogeneous membrane: a concentration-gradient-driven energy harvesting device. Journal of the American Chemical Society. 2015;137(46):14765-72.
41. Zhang Z, Wen L, Jiang L. Bioinspired smart asymmetric nanochannel membranes. Chemical Society Reviews. 2018;47(2):322-56.
42. Bretscher MS. Membrane Structure: Some General Principles: Membranes are asymmetric lipid bilayers in which cytoplasmically synthesized proteins are· dissolved. Science. 1973;181(4100):622-9.
43. Lorent J, Levental KR, Ganesan L, Rivera-Longsworth G, Sezgin E, Doktorova M, et al. Plasma membranes are asymmetric in lipid unsaturation, packing and protein shape. Nature chemical biology. 2020;16(6):644-52.
44. Doktorova M, Symons JL, Levental I. Structural and functional consequences of reversible lipid asymmetry in living membranes. Nature chemical biology. 2020;16(12):1321-30.
45. London E. Membrane structure–function insights from asymmetric lipid vesicles. Accounts of chemical research. 2019;52(8):2382-91.
46. Kamiya K, Kawano R, Osaki T, Akiyoshi K, Takeuchi S. Cell-sized asymmetric lipid vesicles facilitate the investigation of asymmetric membranes. Nature chemistry. 2016;8(9):881-9.
47. Low W-Y, Thong S, Chng S-S. ATP disrupts lipid-binding equilibrium to drive retrograde transport critical for bacterial outer membrane asymmetry. Proceedings of the National Academy of Sciences. 2021;118(50):e2110055118.
48. Powers MJ, Trent MS. Phospholipid retention in the absence of asymmetry strengthens the outer membrane permeability barrier to last-resort antibiotics. Proceedings of the National Academy of Sciences. 2018;115(36):E8518-E27.
49. Henderson JC, Zimmerman SM, Crofts AA, Boll JM, Kuhns LG, Herrera CM, et al. The power of asymmetry: architecture and assembly of the Gram-negative outer membrane lipid bilayer. Annual review of microbiology. 2016;70(1):255-78.
50. Guest RL, Lee MJ, Wang W, Silhavy TJ. A periplasmic phospholipase that maintains outer membrane lipid asymmetry in Pseudomonas aeruginosa. Proceedings of the National Academy of Sciences. 2023;120(30):e2302546120.
51. Arriaga LR, Datta SS, Kim SH, Amstad E, Kodger TE, Monroy F, et al. Ultrathin shell double emulsion templated giant unilamellar lipid vesicles with controlled microdomain formation. small. 2014;10(5):950-6.

52. do Nascimento DbF, Arriaga LR, Eggersdorfer M, Ziblat R, Marques MdFV, Reynaud F, et al. Microfluidic fabrication of pluronic vesicles with controlled permeability. Langmuir. 2016;32(21):5350-5.
53. Perez A, Hernández R, Velasco D, Voicu D, Mijangos C. Poly (lactic-co-glycolic acid) particles prepared by microfluidics and conventional methods. Modulated particle size and rheology. Journal of Colloid and Interface Science. 2015;441:90-7.
54. Hwang S-J, Moon S-K, Kim SE, Kim JH, Choi S-W. Production of uniform emulsion droplets using a simple fluidic device with a peristaltic pump. Macromolecular Research. 2014;22:557-61.
55. Shum HC, Santanach-Carreras E, Kim J-W, Ehrlicher A, Bibette J, Weitz DA. Dewetting-induced membrane formation by adhesion of amphiphile-laden interfaces. Journal of the American Chemical Society. 2011;133(12):4420-6.
56. Hayward RC, Utada AS, Dan N, Weitz DA. Dewetting instability during the formation of polymersomes from block-copolymer-stabilized double emulsions. Langmuir. 2006;22(10):4457-61.
57. Utada A, Chu L-Y, Fernandez-Nieves A, Link D, Holtze C, Weitz D. Dripping, jetting, drops, and wetting: The magic of microfluidics. Mrs Bulletin. 2007;32(9):702-8.
58. Shum HC, Kim J-W, Weitz DA. Microfluidic fabrication of monodisperse biocompatible and biodegradable polymersomes with controlled permeability. Journal of the American Chemical Society. 2008;130(29):9543-9.
59. Peschka D, Haefner S, Marquant L, Jacobs K, Münch A, Wagner B. Signatures of slip in dewetting polymer films. Proceedings of the National Academy of Sciences. 2019;116(19):9275-84.
60. Kim S-H, Kim JW, Kim D-H, Han S-H, Weitz DA. Enhanced-throughput production of polymersomes using a parallelized capillary microfluidic device. Microfluidics and nanofluidics. 2013;14:509-14.
61. Amstad E, Kim SH, Weitz DA. Photo-and thermoresponsive polymersomes for triggered release. Angewandte Chemie International Edition. 2012;51(50):12499-503.
62. Shum HC, Thiele J, Kim S-H. Microfluidic fabrication of vesicles. Advances in transport phenomena 2011. 2013:1-28.
63. Kim S-H, Shum HC, Kim JW, Cho J-C, Weitz DA. Multiple polymersomes for programmed release of multiple components. Journal of the American Chemical Society. 2011;133(38):15165-71.
64. Qian W, Song X, Feng C, Xu P, Jiang X, Li Y, et al. Construction of PEG-based amphiphilic brush polymers bearing hydrophobic poly (lactic acid) side chains via successive RAFT polymerization and ROP. Polymer Chemistry. 2016;7(19):3300-10.
65. Gambin Y, Lopez-Esparza R, Reffay M, Sierecki E, Gov N, Genest M, et al. Lateral mobility of proteins in liquid membranes revisited. Proceedings of the National Academy of Sciences. 2006;103(7):2098-102.
66. Espinosa G, López-Montero I, Monroy F, Langevin D. Shear rheology of lipid monolayers and insights on membrane fluidity. Proceedings of the National Academy of Sciences. 2011;108(15):6008-13.
67. Göpfrich K, Haller B, Staufer O, Dreher Y, Mersdorf U, Platzman I, et al. One-pot assembly of complex giant unilamellar vesicle-based synthetic cells. ACS synthetic biology. 2019;8(5):937-47.
68. Nirasay S, Badia A, Leclair G, Claverie JP, Marcotte I. Polydopamine-supported lipid bilayers. Materials. 2012;5(12):2621-36.
69. Przybylo M, Sýkora J, Humpolickova J, Benda A, Zan A, Hof M. Lipid diffusion in giant unilamellar vesicles is more than 2 times faster than in supported phospholipid bilayers under identical conditions. Langmuir. 2006;22(22):9096-9.
70. Schaich M, Sobota D, Sleath H, Cama J, Keyser UF. Characterization of lipid composition and diffusivity in OLA generated vesicles. Biochimica et Biophysica Acta (BBA)-Biomembranes. 2020;1862(9):183359.
71. Angeletti C, Nichols JW. Dithionite quenching rate measurement of the inside− outside membrane bilayer distribution of 7-nitrobenz-2-oxa-1, 3-diazol-4-yl-labeled phospholipids. Biochemistry. 1998;37(43):15114-9.
72. Gomišček G, Arrigler V, Gros M, Zupančič M, Svetina S. Asymmetrical labeling of giant phospholipid vesicles. Pflügers Archiv-European Journal of Physiology. 2000;440:R051-R2.
73. McIntyre JC, Sleight RG. Fluorescence assay for phospholipid membrane asymmetry. Biochemistry. 1991;30(51):11819-27.
74. Leomil FS, Stephan M, Pramanik S, Riske KA, Dimova R. Bilayer charge asymmetry and oil residues destabilize membranes upon poration. Langmuir. 2024;40(9):4719-31.
75. Henriksen JR, Ipsen JH. Measurement of membrane elasticity by micro-pipette aspiration. The European physical journal E. 2004;14:149-67.
76. Vaziri A, Mofrad MRK. Mechanics and deformation of the nucleus in micropipette aspiration experiment. Journal of biomechanics. 2007;40(9):2053-62.
77. Rawicz W, Olbrich KC, McIntosh T, Needham D, Evans E. Effect of chain length and unsaturation on elasticity of lipid bilayers. Biophysical journal. 2000;79(1):328-39.

78. Dasgupta R, Miettinen MS, Fricke N, Lipowsky R, Dimova R. The glycolipid GM1 reshapes asymmetric biomembranes and giant vesicles by curvature generation. Proceedings of the National Academy of Sciences. 2018;115(22):5756-61.
79. Guevorkian K, Gonzalez-Rodriguez D, Carlier C, Dufour S, Brochard-Wyart F. Mechanosensitive shivering of model tissues under controlled aspiration. Proceedings of the National Academy of Sciences. 2011;108(33):13387-92.
80. Di Cerbo A, Rubino V, Morelli F, Ruggiero G, Landi R, Guidetti G, et al. Mechanical phenotyping of K562 cells by the Micropipette Aspiration Technique allows identifying mechanical changes induced by drugs. Scientific reports. 2018;8(1):1219.
81. Brugués J, Maugis B, Casademunt J, Nassoy P, Amblard F, Sens P. Dynamical organization of the cytoskeletal cortex probed by micropipette aspiration. Proceedings of the National Academy of Sciences. 2010;107(35):15415-20.
82. Guevorkian K, Maître J-L. Micropipette aspiration: A unique tool for exploring cell and tissue mechanics in vivo.  Methods in cell biology. 139: Elsevier; 2017. p. 187-201.
83. Lee LM, Liu AP. The application of micropipette aspiration in molecular mechanics of single cells. Journal of nanotechnology in engineering and medicine. 2014;5(4):040902.
84. González-Bermúdez B, Guinea GV, Plaza GR. Advances in micropipette aspiration: applications in cell biomechanics, models, and extended studies. Biophysical Journal. 2019;116(4):587-94.
85. Hochmuth RM. Micropipette aspiration of living cells. Journal of biomechanics. 2000;33(1):15-22.
86. Hsu C-P, Aretz J, Hordeichyk A, Fässler R, Bausch AR. Surface-induced phase separation of reconstituted nascent integrin clusters on lipid membranes. Proceedings of the National Academy of Sciences. 2023;120(31):e2301881120.
87. Dimova R. Giant vesicles and their use in assays for assessing membrane phase state, curvature, mechanics, and electrical properties. Annual review of biophysics. 2019;48(1):93-119.
88. Pincet F, Adrien V, Yang R, Delacotte J, Rothman JE, Urbach W, et al. FRAP to characterize molecular diffusion and interaction in various membrane environments. PloS one. 2016;11(7):e0158457.
89. Kang M, Day CA, Kenworthy AK, DiBenedetto E. Simplified equation to extract diffusion coefficients from confocal FRAP data. Traffic. 2012;13(12):1589-600.
90. Xiao Q, Yadavalli SS, Zhang S, Sherman SE, Fiorin E, Da Silva L, et al. Bioactive cell-like hybrids coassembled from (glyco) dendrimersomes with bacterial membranes. Proceedings of the National Academy of Sciences. 2016;113(9):E1134-E41.
91. Krywko-Cendrowska A, Di Leone S, Bina M, Yorulmaz-Avsar S, Palivan CG, Meier W. Recent advances in hybrid biomimetic polymer-based films: from assembly to applications. Polymers. 2020;12(5):1003.
92. Lu Y, Allegri G, Huskens J. Vesicle-based artificial cells: materials, construction methods and applications. Materials horizons. 2022;9(3):892-907.
93. Trantidou T, Dekker L, Polizzi K, Ces O, Elani Y. Functionalizing cell-mimetic giant vesicles with encapsulated bacterial biosensors. Interface Focus. 2018;8(5):20180024.
94. Mohammadi M, Taghavi S, Abnous K, Taghdisi SM, Ramezani M, Alibolandi M. Hybrid vesicular drug delivery systems for cancer therapeutics. Advanced Functional Materials. 2018;28(36):1802136.
95. Pocanschi CL, Dahmane T, Gohon Y, Rappaport F, Apell H-J, Kleinschmidt JH, et al. Amphipathic polymers: tools to fold integral membrane proteins to their active form. Biochemistry. 2006;45(47):13954-61.
96. Manafirad A, Menendez CA, Perez-Lemus GR, Thayumanavan S, de Pablo JJ, Dinsmore AD. Structural and mechanical response of two-component photoswitchable lipid bilayer vesicles. Langmuir. 2023;39(45):15932-41.

**Figure Legends**

Figure 1. (A) (top) illustration and image (bottom) of the microfluidic device (B) an exemplar sample, with C.V analysis by image processing analysis from MATLAB. The images show contour detection and calculates the C.V to be 5.2%(C) Schematic and (D) optical image of double emulsion transformation into vesicles. (E)Fluorescence images of vesicles labelled with 3 wt% 18:1 Liss Rhod PE, the red lipophilic dye, and 3 wt% FITC-PEG-b-PLA, the green polymer dye. The lipids and polymers are distributed throughout the membrane. (F) Fluorescence images of vesicles labelled with 3 wt% 18:1 Liss Rhod PE in red, the DOPC affinity dye, and 3 wt% naphtopyrene, the DPPC affinity dye in blue. The lipids, DOPC and DPPC, form a lipid rich region (red and blue), separates from the polymer rich region, which is non-fluorescent.

Figure 2. (A) (Left): Schematic and optical image of the microfluidic device. (B) Optical images showing uniform emulsions with CV approximately at 5.5%. (C)Schematic and (D) optical images illustrating the transformation of triple emulsions into asymmetric vesicles, as the oil, which appears dark, leaves the vesicle membranes. (E) Fluorescence imaging reveals lipid labelled by rhodamine and polymer labelled by FITC are both incorporated into the membranes. (F-H) Quenching of fluorescence in asymmetric vesicles. In the asymmetric vesicle with lipid outer, the outer lipid leaflet is labeled with 3 wt% NBD-DOPE, a fluorescent dye. Upon addition of dithionite, the outer leaflet is exposed to the quencher. In the asymmetric vesicle with lipid inner, the inner lipid leaflet is labeled with 3 wt% NBD-DOPE and is shielded from the quencher. Its fluorescence is only slightly quenched compared to the drop observed in (F), indicating that most lipids reside in the inner leaflet, thereby confirming membrane asymmetry.

Figure 3. Mechanical properties vesicles (A) (left) schematic illustration and optical images of a vesicle being aspirated in a micropipette by an incremental suction pressure, $\Delta P$, until it ruptures. We detect the contours of the vesicle both inside and outside of the pipette, which allow us to determine, $R_s$ and $R_p$, radii of the vesicle outside and inside of the pipette respectively, and $L$, the length of the vesicle being sucked into the pipette, which are used to calculate the surface area of the vesicle in the suppplement. To calculate the surface tension, $\tau$, we use $\tau = \frac{\Delta P}{2(\frac{1}{R_p} - \frac{1}{R_s})}$. (B) table of formulations. F1 to F3 are sym vesicles containing increasing polymer contents; F4-F5 are two asy vesicles with different leaflet configuration and asymmetric degree; F6 is a pure polymersome group (C) Membrane tension versus areal strain curves. For symmetric vesicles, F1 to F3, increasing polymer content increases the stretching modulus of vesicles. For the asymmetric vesicle with lipid inner and polymer outer, F5, the vesicles are both stiff and stretchable as compared to sym groups, but stiffer. (D) rupture of all types of vesicles reveals their liquid or solid like behaviors. F1, F2, and F4, which all contain amounts of lipids exceeding 40%, behave more like liquid vesicles without wrinkles at rupture. F3, F5, and F6, which either have more amounts of polymers, or likely a connected polymer leaflet, behave more like solid vesicles with wrinkles at rupture. (E), (F) Statistical summary across all vesicle types. The results show that asymmetric vesicles with high asymmetry have approximately half the stretching modulus but greater stretchability, as compared to polymersomes, resulting in its highest toughness among all vesicle types.

Figure 4. FRAP measurements of lipid diffusivity in vesicles (A) (Top) schematic and (bottom) optical images showing a laser photobleaching a circular area of a fluorescently labeled vesicle and the fluorescence recovers due to the diffusion of lipid molecules. (B) Table of all formulations (C) Fluorescence recovery in the bleached area with radius of approximately 10 $\mu$m. The fluorescence intensity in the bleached area, $I(t)$, is normalized to an intensity in the control area, which is of the same radius but unbleached, (D) log intensity of fluorescence recovery curves shows normal diffusion of lipids. The log plots all behave linearly, indicating the lipids diffuse normally, allowing us to fit the recovery intensity to a simple equation, $I(t) = A\left(1 - \exp\left(-\frac{t}{\tau_D}\right)\right)$, We obtain the lifetime of recovery,$\tau_D$, and half time of recovery, $\tau_{1/2} = -(\ln 0.5)\tau_D$. (E) Process for bleaching vesicles with varied bleached radii (F) Plot of the square of the radius of the bleached spot, $r^2$, versus the half time, $\tau$. With increasing polymer content in symmetric vesicles, the linear slope decreases as shown by F1 to F3. With increasing asymmetry degree, the slope decreases even more as shown by F4 and F5. (G) The diffusion coefficients of the vesicles across various types are calculated from the slopes of the lines and represented in a bar plot. The highly asymmetric vesicles in V5 have much slower lipid diffusion coefficient than the symmetric vesicles.

(A) microfluidic process for producing double emulsion -> (B) one example of symmetric hybrid vesicles

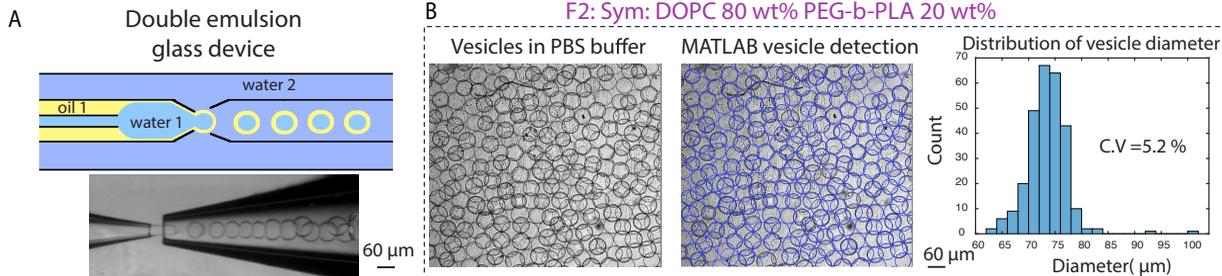

(C) process of symmetric vesicle formation   ->   (D) one example of symmetric hybrid vesicles

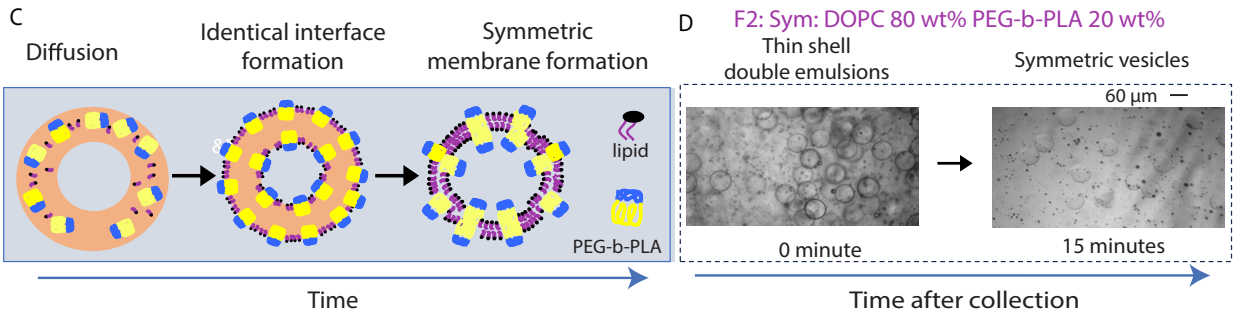

(E) homogeneou distribution of DOPC and polymer ->(F) phase separation in DPPC and polymer membrane

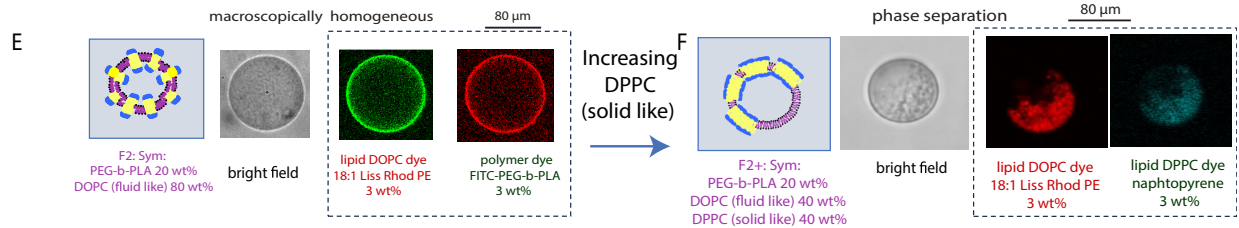

(A) schematic for triple emulsion device -> (B) one example of asymmetric vesicles with lipid in and polymer out

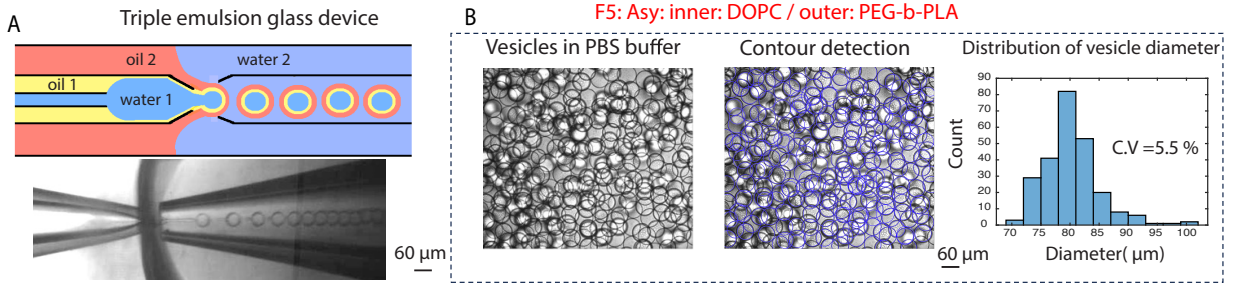

(C) schematic of dewetting process -> (D) imaging of dewetting -> (E) fluorescence of formed vesicles

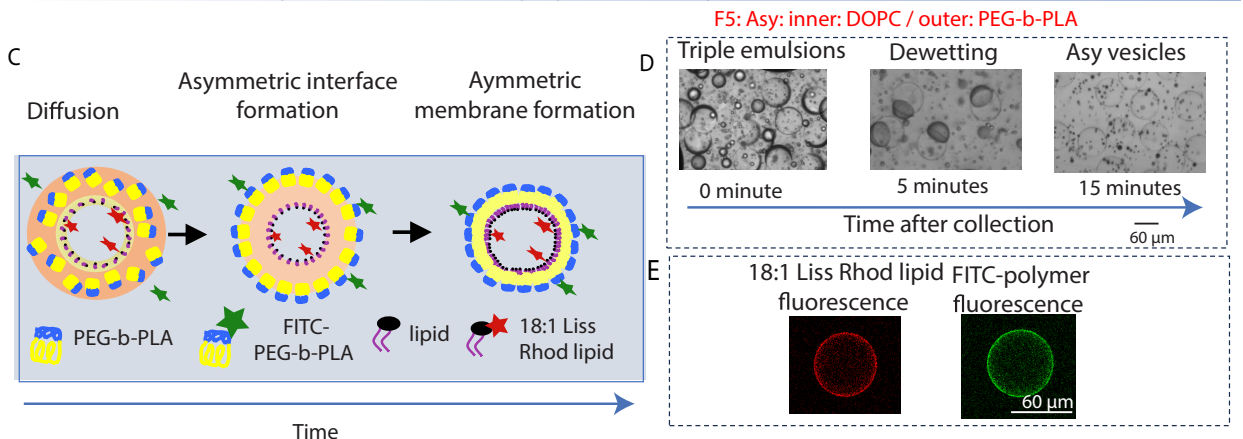

(F-H) Characterizing asymmetry degree in asy vesicles with polymer inner and lipid outer (F4) and in asy vesicles with polymer outer and lipid inner (F5).

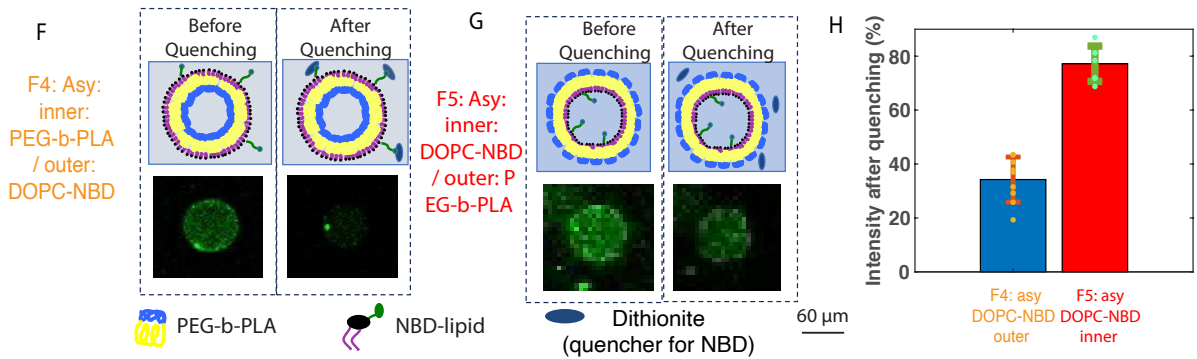

## (A), (B), (C) micropippette aspiration setup and measurements for F1-F6

## (D) ruptures of vesicles at the end of aspiration

## (E), (F) statistics of mechanical properties for F1-F6

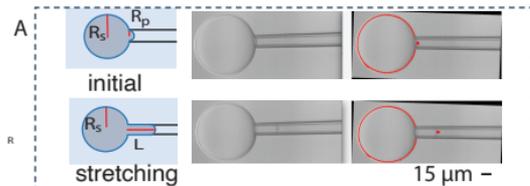

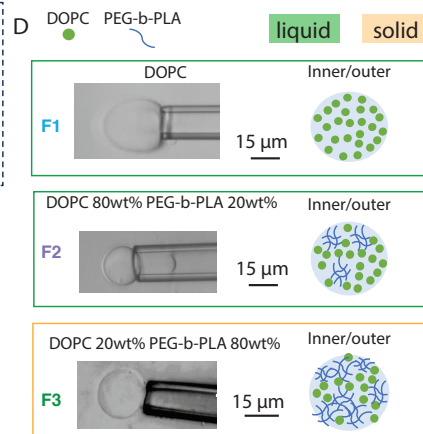

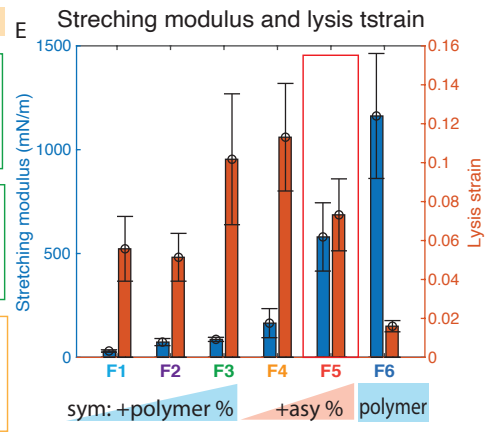

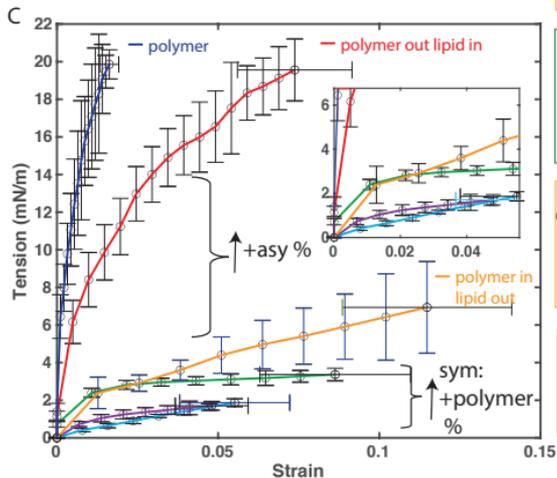

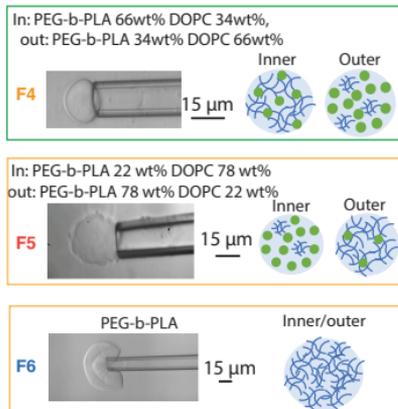

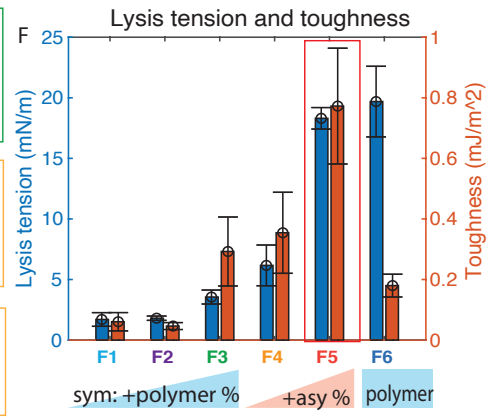

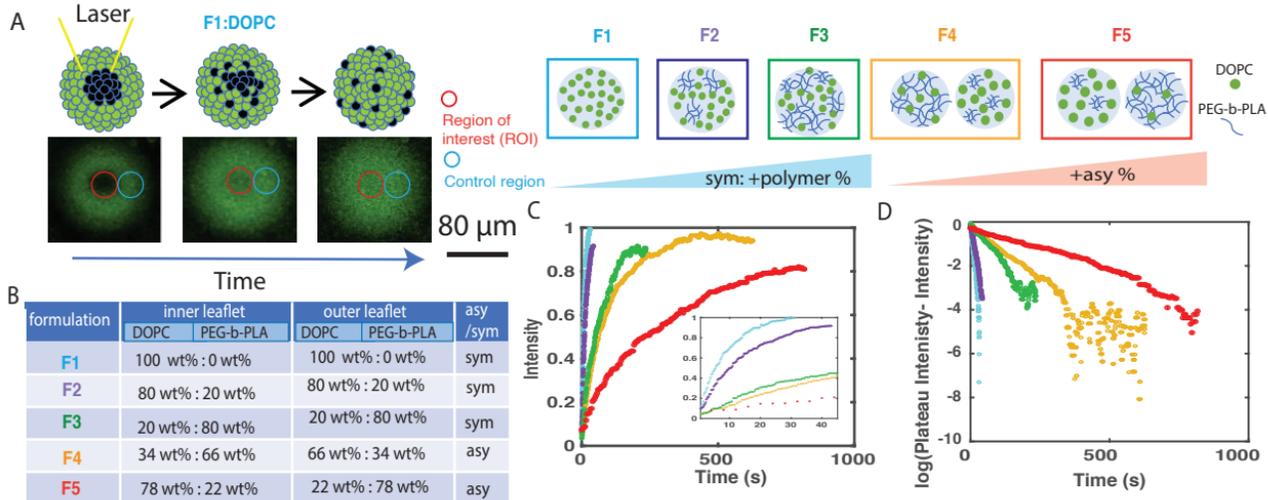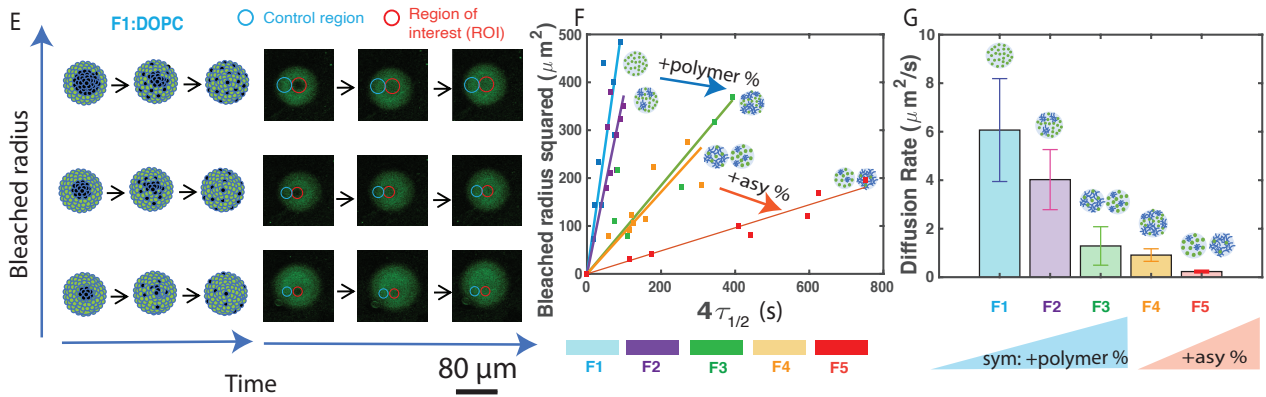